\documentclass[12pt]{article}
\usepackage[dvips]{graphicx}
\usepackage{latexsym}
\usepackage{amssymb}

\newcommand{\CP}{\mathbb{CP}}

\newcommand{\R}{\mathbb{R}}

\renewcommand{\d}{\mathrm{d}}

\def\be{\begin{equation}}
\def\ee{\end{equation}}

\def\om{\omega}

\def\p{\partial}

\def\l{\lambda}

\newtheorem{theo}{Theorem}[section] 
\newtheorem{prop}[theo]{Proposition}

\begin{document}
\pagestyle{plain}

\title{\vskip -70pt
\begin{flushright}
{\normalsize DAMTP-2006--3} \\
\end{flushright}
\vskip 80pt
{\bf Einstein--Maxwell--Dilaton metrics from three--dimensional Einstein--Weyl
structures. }
\vskip 20pt}

\author{Maciej Dunajski\thanks{email M.Dunajski@damtp.cam.ac.uk}\\[15pt]
{\sl Department of Applied Mathematics and Theoretical Physics} \\[5pt]
{\sl University of Cambridge} \\[5pt]
{\sl Wilberforce Road, Cambridge CB3 0WA, UK} 
}
\date{} 
\maketitle
\begin{abstract}
A class of time dependent solutions to $(3+1)$ Einstein--Maxwell-dilaton theory
with attractive electric force is found from Einstein--Weyl structures
in (2+1) dimensions corresponding to  dispersionless Kadomtsev--Petviashvili 
and $SU(\infty)$ Toda equations. These solutions are obtained from time--like
Kaluza--Klein reductions of $(3+2)$ solitons.
\end{abstract}
\newpage
\section{Introduction}
\setcounter{equation}{0}

Singularity theorems of Hawking and Penrose assert that mild energy conditions
imposed on the energy momentum tensor result in a gravitational collapse 
to a singularity. While the final singular state of a collapsing star is 
inevitable, not much is known about the dynamical mechanisms leading to the 
formations of the singularities. The relevant time dependent exact solutions
to Einstein equations are unknown, and the numerical considerations are made
difficult by the Birkhoff theorem which says that any spherically symmetric
vacuum solution is static, which implies that the metric is Schwarzchild.
To make progress one would need to draw conclusions from
numerical evolution of  a non-spherically
symmetric initial data which is considerably more difficult.

One way to overcome these difficulties is to introduce matter fields to Einstein equations which allows the  study of time evolution, while maintaining the 
spherical symmetry. The simplest choice corresponds to the massless scalar 
field. Christodoulou has given a complete analysis of this situation 
\cite{Ch1}.
Interestingly enough his analysis revealed that certain data on a future null
cone centred at the origin can evolve into a solution with naked singularities
(i.e. singularities not hidden inside an event horizon) in contrary with
the Cosmic Censorship Hypothesis (CCH). This is a mild violation of the CCH,
as the initial data is given on a high co-dimension surface inside the null 
cone, and is unstable in a sense that any data away from this surface does not
evolve into naked singularities. 

Another way to evade Birkhoff's theorem is to go to more than four
space--time dimensions. This was recently done in \cite{Bizon} where
the following ansatz was made for a (4+1) metric\footnote{In this paper 
$g_{(r,s)}$ denotes a real pseudo--Riemannian metric of signature $(r, s)$.} 
\be
\label{bizon}
g_{(4,1)}=-C e^{-2\delta}\d t^2 +C^{-1}\d r^2+\frac{1}{4}r^2
(e^{2B}(\sigma_1^2+\sigma_2^2)+e^{-4B}\sigma_3^2).
\ee
Here $\sigma_i, i=1,2,3$ are the left invariant one forms on $SU(2)$ satisfying the standard Maurer--Cartan relations, and the functions $B, C, \delta$ 
depend on $(r, t)$. The authors of \cite{Bizon} have numerically studied the PDEs
for these functions resulting from the Ricci-flatness of (\ref{bizon}), and
have shown that a (4+1) dimensional Schwarzchild black hole 
\[
C=1-\frac{\mbox{const}}{\rho^2} ,\qquad B=0, \qquad \delta=0
\]
is formed for a large initial data. 
This lead to an explicit numerical profile of settling down to a singularity.

Gibbons \cite{gib_talk} 
has pointed out that the gravitational collapse of (\ref{bizon})
is not as inevitable as it may seem from the numerical analysis. A 
spherically symmetric star in (4+1) dimensions can also settle into a
soliton, i.e. a non-singular topologically stable solution
of the field equations. The simples case is the Kaluza--Klein
monopole of Gross--Perry and Sorkin \cite{GP, Sor}, where the five--dimensional
metric takes the form
\be
\label{KK_1}
g_{(4,1)}=-\d t^2+g_{TAUB-NUT},
\ee
where $g_{TAUB-NUT}$ is the simplest asymptotically locally flat (ALF)
four dimensional gravitational instanton. Any ALF gravitational instanton would
do, so a mild, triaxial, generalisation of the ansatz (\ref{bizon}) admits
a static soliton of the form (\ref{KK_1}) with $g_{TAUB-NUT}$ replaced
by the Atiyah--Hitchin gravitational instanton \cite{AH}, where the 
complex structures are rotated by the $SU(2)$ action. Using the 
gravitational instantons  
can even lead to explicit time dependent solutions. As pointed out in 
\cite{GLP05}
exploiting the scaling symmetry in the $A_k$ ALF multi--instanton leads
to the explicit solution of the 4+1 Einstein equations
\be
\label{G_H}
g_{(4,1)}=-\d t^2+V\;h_{flat}+V^{-1}(\d\theta+A)^2,
\ee
where
\be
\label{time_dep}
V=t+\sum_{i=1}^k\frac{m}{|{\bf r}-{\bf r_k}|}
\ee
is a solution to the three--dimensional Laplace equation $\triangle V=0$
depending on a parameter $t$ 
and the one--form $A$ satisfies 
the monopole equation $\d A=*_3 \d V$, where $*_3$ is
the duality operator of the flat Euclidean metric $h_{flat}$ 
in three dimensions.
Here $m$ is a constant mass parameter, and ${\bf r}_1, ..., {\bf r}_k$
are positions of fixed points in $\R^3$. The metric appears singular at these
points, but in fact it is not if $\theta$ is taken to be periodic, and 
the constant $m$ equals to  half of the period. 
In particular choosing $k=1$ and
${\bf r}_1={\bf 0}$ leads to a time--dependent generalisation of the 
Taub--NUT Kaluza--Klein monopole
metric which is an explicit solution to the Einstein equations imposed
on (\ref{bizon}). 

From the (3+1) dimensional perspective the solutions discussed so far
give rise to solutions of Einstein--Maxwell theory with a dilaton. This is
the standard Kaluza--Klein reduction where the fifths dimension compactifies
to a circle of a small radius. This corresponds to $\theta$ in (\ref{G_H}) 
being periodic. If the radius is sufficiently 
small then low energy experiments will average
over the fifths dimension thus leading to an effective four--dimensional theory
with the Maxwell potential given by $A$, and the dilaton given by
$-(\sqrt{3}/4)\log{V}$. 
One even gets one time dependent solution (\ref{time_dep}), 
but this seems
to be an isolated case. 

 The purpose of this paper is to point out (Proposition \ref{proposition})
that large families of explicit time dependent solutions can be found
in the $(3+1)$ dimensional theory. They will come from  time--like 
Kaluza--Klein reductions of pure Einstein equations in $(3+2)$ dimensions.
The five--dimensional metrics $g_{(3, 2)} $ are given by 
four--dimensional Ricci--flat metrics $g_{(2,2)}$ of signature $(2, 2)$  
\be
\label{G32}
g_{(3,2)}=\d z^2 +g_{(2,2)}. 
\ee
To make the Kaluza--Klein reduction possible the metric $g_{(2, 2)}$ must admit a
Killing vector, and so be of the form
\be
\label{g22}
g_{(2, 2)}=Vh_{(2, 1)}-V^{-1}(\d \theta+A)^2
\ee
where $h_{(2,1)}$ is a metric of signature $(2, 1)$ on the three--dimensional 
space of orbits of the Killing vector $\p/\p \theta$, and
$(V, A)$ is a function and a one--form on this space. 
The physical metric $G_{\mu\nu}$ of signature $(3, 1)$ is then 
given from (\ref{G32}) by
\be
\label{g31}
g_{(3, 2)}=\exp{(-2\Phi/\sqrt{3})}G_{\mu\nu}\d x^{\mu}\d x^{\nu}
-\exp{(4\Phi/\sqrt{3})}(\d\theta+A)^2.
\ee
This metric has a space--like Killing vector $\p/\p z$ but is not stationary
if $h_{(2, 1)}$ does not admit a time--like Killing vector. In the next
section we shall construct explicit examples taking $g_{(2, 2)}$ to
be a Ricci--flat anti--self--dual metric with symmetry. This will imply
that $h_{(2, 1)}$ is a part of the so called Einstein--Weyl 
structure \cite{Hi82,JT85}, and can in principle be found explicitly by twistor
methods. 

If the reader objects to using the metric $g_{(3,2)}$ of signature 
$(3, 2)$ as non-physical, he should regard it as a mathematical trick 
for producing interesting Lagrangians in four dimensions. A more serious
objection comes from performing the K-K reduction along the time--like 
symmetry. This will result in a change of the relative sign between the
Ricci scalar and the Maxwell term in the four dimensional effective 
Lagrangain.  The charges in the resulting electro--vacuum solutions 
will therefore attract rather than repel, thus ruling out the 
extremality condition. The formalism
is nevertheless well adopted to studying the Einstein anti--Maxwell theory 
\cite{GR}.
\section{Anti--Self--Dual Ricci--flat $(2, 2)$ metrics with symmetry}
Consider a $(2, 2)$ signature metric $g_{(2,2)}$ on a four--dimensional 
manifold $M$.  The construction outlined 
in the introduction demands that it is Ricci flat, but to find explicit 
examples we shall also assume that its curvature (when viewed as a two--form)
is anti--self--dual. In the $(2, 2)$ signature the spin group $Spin(2, 2)$ 
decomposes as a product of two independent copies of $SL(2, \R)$, and
the representation space of the spin group splits up into a direct sum of
two real two dimensional vector spaces $S_+$ and $S_-$. In the ASD Ricci flat
case the curvature of the spin connection on $S_+$ is zero, 
and the holonomy effectively reduces to 
$SL(2, \R)$. From the mathematical point of view  $g_{(2,2)}$ is a
pseudo-Riemannian analog of a four dimensional hyper--Kahler structure.
The endomorphisms of the tangent bundle associated to three Kahler structures
satisfy the algebra of pseudo--quaternions.

Any 
ASD Ricci--flat $(2, 2)$ metric with a non-null symmetry 
is of the form (\ref{g22}) and, by the Jones--Tod correspondence \cite{JT85},
the ASD vacuum  equations reduce down to
the Lorentzian Einstein--Weyl equations in three--dimensional space of 
orbits $W$ of the Killing vector $K=\p/\p \theta$ (if the symmetry is null
the ASD vacuum equations linearise  \cite{DW06}). 
This means that there
exists a torsion--free connection $D$ on $W$ such that the null 
geodesics of a 
conformal structure $[h]$ defined by $h_{(2,1)}$ are also geodesic of 
this connection. 
This compatibility condition implies the existence
of a one--form $\om$ on $W$ such that
\[
D h_{(2, 1)}=\om\otimes h_{(2, 1)}.
\] 
If 
we  change this representative by $h\rightarrow \psi^2 h$, 
then $\om\rightarrow \om +2\d \ln{\psi}$, where $\psi$ is a 
non-vanishing function on $W$.

The pair $(D, [h])$ satisfies the conformally invariant Einstein--Weyl 
equations which assert that the symmetrized Ricci tensor of $D$ 
is proportional to $h_{(2, 1)}$. 
One can regard $h_{(2, 1)}$ and $\om$ as the unknowns in these equations.
Once they have been  found, the covariant differentiation w.r.t $D$ is given by
\[
D\chi=\nabla \chi-\frac{1}{2}(\chi\otimes \om+(1-m)\om\otimes V-h_{(2, 1)}(\om,\chi)h_{(2, 1)}),
\] 
where $\chi$ is a one--form of conformal weight $m$, and $\nabla$ is 
the Levi--Civita connection of $h$.

Given the Einstein--Weyl structure which arises
from and ASD vacuum structure, the metric $g_{(2, 2)}$ is given by 
solutions to the generalised monopole equation \cite{JT85}
\be
\label{monopole}
*_h(\d V+\frac{1}{2}\om V)=\d A,
\ee
where $*_h$ is the duality operator in three dimensions corresponding to 
$h_{(2, 1)}$,
and the unknowns are the function $V$ and the one--form $A$ on $W$. 
The arbitrary solution to the generalised monopole equation would lead
to conformally ASD metric $g_{(2, 2)}$ which is not necessarily vacuum.
One must therefore select a special class of solutions. This problem 
has been extensively analysed, and is known to lead to three possibilities
depending
on $(\nabla K)_+$, the self dual derivative of the Killing vector:
\begin{itemize}
\item
If $(\nabla K)_+$ is zero then $K$ preserves the self--dual two
forms on $M$, and the moment map coordinates can be chosen. In this case
If $h_{(2, 1)}$ is flat,
\be
\label{22GH}
g_{(2, 2)}=V h_{flat}-V^{-1}(\d \theta+A)^2
\ee
and $(V, A)$ satisfy the 2+1 monopole equation $*\d V=\d A$.
\item
If $(\nabla K)_+$ is a simple two form (i.e.$(\nabla K)_+ =\d p\wedge\d q$
for some functions $(p, q)$ on $M$) then there exist local coordinates such 
that \cite{DMT}
\be
\label{dkpmet}
h_{(2,1)}=\d y^2-4\d x\d t-4 u\d t^2,  
\ee
and 
\be
\label{HCdKP}
g_{(2, 2)}=\frac{u_x}{2}h_{(2,1)}
-\frac{2}{u_x}(\d \theta-\frac{u_x\d y}{2}-u_y\d t)^2,
\ee
where the function $u=u(x, y, t)$ satisfies the dispersionless
Kadomtsev--Petviashvili equation
\be
\label{dKP}
(u_t-uu_x)_x=u_{yy}.
\ee
Note that $V=u_x/2\neq 0$ for (\ref{HCdKP}) to be well defined. 
The flat metric $g_{(2,2)}$ corresponds to $u=-x/t$.
\item
Finally if $(\nabla K)_+\wedge (\nabla K)_+\neq 0$ there exist local 
coordinates such that
\be
\label{Todamet}
{h_{(2, 1)}}^{\pm} =e^u(\d x^2\pm\d y^2)\mp\d t^2,
\ee
and
\be
\label{Toda4}
g_{(2, 2)}=\frac{u_t}{2}h_{(2,1)} -\frac{2}{u_t}(\d\theta+A)^2,
\ee
where the one--form $A$ on $W$ is a solution to the linear equation
(\ref{monopole}) with $V=u_t/2$ and $\omega=2u_t\d t$. 
The ASD vacuum  equations reduce \cite{W90,L91,T95} 
to the $SU(\infty)$ Toda equation 
\be
\label{BFP}
u_{xx}\pm u_{yy}\mp {(e^u)}_{tt}=0.
\ee
The $\pm$ sign in (\ref{BFP}) correspond to two different Lorentzian slices 
$h_{(2, 1)}$.

\end{itemize}
The reader should note that while the case (\ref{G_H}) and (\ref{Todamet}) are 
indefinite analogues of positive--definite metrics, the case
(\ref{dkpmet}) is genuinely pseudo--Reimannian, as there are no null two--forms
in the Riemannian case.

\section{Kaluza--Klein reduction}
Let $h_{(2,1)}$ be an Einstein--Weyl given by one of 
(\ref{dkpmet}, \ref{Todamet}) or $h_{flat}$,
and let $(V, A)$ be the corresponding solution to the generalised 
monopole equation
giving rise  to a vacuum metric (\ref{g22}). Consider the $(3+2)$ 
dimensional metric $g_{(3, 2)}$ given by (\ref{G32}). It is obviously 
Ricci--flat, and it
admits
two commuting Killing vectors $\p/\p z$ and $\p/\p \theta$. We 
preform the Kaluza-Klein reduction with respect to the time like vector
 $\p/\p\theta$. The four dimensional theory is invariant under the general coordinate transformations independent of $\theta$. The translation of the fiber
coordinate $\theta\rightarrow \theta+\Lambda(x^{\mu})$ induces
the $U(1)$ transformations of the Maxwell one--form.
The scaling of 
\[
\theta\longrightarrow \l\theta, 
\qquad [g_{(3,2)}]_{\mu\theta}\longrightarrow \l^{-2}
[g_{(3,2)}]_{\mu\theta},
\qquad 
\]
is spontaneously broken by the Kaluza--Klein 
vacuum, since $\theta$ is a coordinate
on a circle with a fixed radius. The scalar field corresponding to this
symmetry breaking is called the dilaton. It is the usual practise
to  conformally rescale the resulting $(3+1)$ dimensional metric, and the
dilaton so that the multiple of the Ricci scalar of $G_{\mu\nu}$ in the reduced Lagrangian is equal to  $\sqrt{|\det(G_{\mu\nu})|}$.
The corresponding Maxwell field is  $F=\d A$, and the physical metric 
$G_{\mu\nu}$ in 
$(3+1)$ signature is given by (\ref{g31}).
We can summarise our findings in the following proposition
\begin{prop}
\label{proposition}
Let $([h], D, W)$ be an Einstein--Weyl structure in 2+1 dimensions,
and let $V$ be function on $W$ of conformal weight $-1$ 
which is a solution to the generalised monopole equation {\em(\ref{monopole})}
such that the corresponding 
$(++--)$ ASD metric {\em(\ref{g22})} is  Ricci--flat. Then for any 
$h_{(2, 1)}\in [h]$ the 
triple
\[
G=\sqrt{V}h_{(2,1)}+\frac{1}{\sqrt{V}}\d z^2, \qquad
\Phi=-\frac{\sqrt{3}}{4}\log{V}, \qquad F=*_h D V
\]
satisfies the Einstein--Maxwell--Dilaton equations arising from 
the Lagrangian 
density  
\be
\frac{1}{16\pi^2}(R-2(\nabla \Phi)^2+ \frac{1}{4}e^{-2\sqrt{3}\Phi} F_{\mu\nu}F^{\mu\nu}).
\ee
\end{prop}
To understand the unusual sign between the Ricci scalar and the Maxwell term 
in this Lagrangian
notice that making the replacements
\[
\theta\longrightarrow i\theta, \qquad A\longrightarrow iA
\]
would lead to a more usual space--like Kaluza--Klein reduction from
$(4+1)$ to $(3+1)$ dimensions, where the relative sign between
the Ricci scalar and  the Maxwell term is negative.

The negative energy of the Maxwell field has peculiar physical consequences.
No static multi--black hole solutions analogous to the Majumdar--Papapetrou 
extremal black holes can exist, as both gravity and electromagnetism are 
now attractive forces and the cancellations can not take place. Thus the theory
only allows non--extremal black holes which can effectively increase
their masses by radiating photons out! We can also encounter
`tachyonic' solutions invariant under $\R\times SO(2, 1)$.

Proposition (\ref{proposition}) also applies to the ordinary EMD solutions
with positive Maxwell energy if one takes $([h], D, W)$ to be positive 
definite Einstein Weyl structure and preforms the usual space-like 
Kaluza--Klein
reduction of the product metric $g_{(4)}-\d t^2$, where $g_{(4)}$ is 
an ASD Ricci flat metric corresponding to $([h], D, W)$. This will rule out
solutions coming from the dKP equation (\ref{dKP}). There is however
another possible construction which we now outline. Let
\[
h_{(3)}=\d z^2+e^{u}(\d x^2+\d y^2), \qquad \om=2u_z  \d z
\]
be a positive--definite EW structure corresponding to a solution
$u=u(x, y, z)$ to the elliptic $SU(\infty)$ Toda equation $u_{xx}+u_{yy}+(e^u)_{zz}=0$.
If $(V, A)$ is an arbitrary solution to 
the monopole equation (\ref{monopole}) the resulting four--dimensional
Riemannian metric
\[
g_{(4)}= Vh_{(3)}+V^{-1}(\d \theta+A)^2
\]
is scalar--flat and Kahler \cite{L91}. One special solution $V=u_z/2$
makes $g_{(4)}$ Ricci--flat, and the Lorentzian version of this solution
must be used in Proposition (\ref{proposition}). If one instead takes
$V=V_{\Lambda}$, where
\[
V_{\Lambda}=-\frac{1}{3}\Lambda(1-zu_z)
\]
then $g_{(4)}$ is conformal to an Einstein metric $\hat{g}_{(4)}$ with a 
cosmological constant $\Lambda$ \cite{T97}:
\be
\hat{g}_{(4)}=\frac{V_{\Lambda}}{z^2}h_{(3)}+
\frac{1}{z^2V_{\Lambda}}(\d\theta +A)^2.
\ee
Given $\hat{g}_{(4)}$ we construct a `cosmological' vacuum metric
in 4+1 dimensions given by 
\[
g_{(4, 1)}=f(t)^2\hat{g}_{(4)}-\d t^2.
\]
There are several possibilities for $f$: 
Choosing $f(t)=L^{-1}\mbox{cosh}{(Lt)}$
yields a regular metric, and $f(t)=L^{-1}\sin{(Lt)}$ gives
a generalised AdS solution with Big Bang and Big Crunch singularities
at $t=0$ and $t=\pi/L$ respectively. 
The K-K reduction of this metric (now we need to change the relative
sign between the two terms in (\ref{g31})) gives the following solution
to the EMD equations
\begin{eqnarray*}
G_{(3, 1)}&=&\frac{f}{z}\Big(\sqrt{V_{\Lambda}}\Big(\frac{f}{z}\Big)^2h_{(3)}-
\frac{1}{\sqrt{V_{\Lambda}}} \d t^2\Big),\qquad
\Phi=-\frac{\sqrt{3}}{2}\log{\Big({\frac{f}{z\sqrt{V_{\Lambda}}}}\Big)},\\
F&=&-(V_{\Lambda})_x\d y\wedge \d z-(V_{\Lambda})_y\d z\wedge\d x-
(V_{\Lambda}e^u)_z\d x\wedge\d y.
\end{eqnarray*}
\section{Example}
The simplest illustration of Proposition (\ref{proposition})
corresponds to $g_{(2, 2)}$ being an
analytic continuation of a gravitational instanton in Riemannian signature.
We shall examine the (3+1) solution arising from a $(2, 2)$ analog of the 
Taub--Nut gravitational instanton, and emphasise that this example
hints the semi-classical instability of the Einstein--anti Maxwell
theories.

Consider (\ref{22GH}), and take
\be
\label{formula_V}
V=\varepsilon+\frac{m}{\sqrt{x^2+y^2-t^2}}, \qquad \varepsilon=1.
\ee

The NUT singularity is absent in $(3+2)$ dimensions. It is a fixed point of the Killing vector $\p/\p\theta$ which is regular if $\theta$ is periodic.
The resulting metric (3+1) metric (\ref{g31}) written in  cylindrical polar 
coordinates takes the form
\begin{eqnarray}
\label{example1}
\d s^2&=&
\Big(\varepsilon+\frac{m}{\sqrt{\rho^2-t^2}}\Big)^{-1/2}
\d z^2+\Big(\varepsilon+\frac{m}{\sqrt{\rho^2-t^2}}\Big)^{1/2}
(\d \rho^2+\rho^2\d \theta^2-\d t^2),\\
\Phi&=&-\frac{\sqrt{3}}{4}\log{(\varepsilon+\frac{m}{\sqrt{\rho^2-t^2}})}, 
\qquad A=\Big(\varepsilon+\frac{m}{{\sqrt{\rho^2-t^2}}}\Big)^{-1}\d z\nonumber,
\end{eqnarray}
and we can take $z$ to be periodic\footnote{Another possibility is 
to take  $V$ as in (\ref{formula_V}) with
$\varepsilon=z$ which leads to a non-asymptotically flat metric.}.

The initial data on the surface $t=0$ is regular everywhere. 
The rescalling of the  three--metric
\[
g_{(3)}=\Big(\varepsilon+\frac{m}{{\rho}}\Big)^{-1/2}\d z^2+
\Big(\varepsilon+\frac{m}{{\rho}}\Big)^{1/2}
(\d \rho^2+\rho^2\d \theta^2)
\]
is regular at $\rho=0$ if $\theta$ is periodic with a period $4\pi$.
This can be seen by setting $\rho=\hat{\rho}^2/m$, so
that around $\hat{\rho}=0$
\[
g_{(3)}\sim \d z^2+4\Big(\d \hat{\rho}^2+\hat{\rho}^2\;\d \Big(\frac{\theta}{2}\Big)^2\Big).
\]
The cylindrical 
mass of $g_{(3)}$ can be defined as a deficit angle at infinity, which
is proportional to the integral of the Gaussian curvature of the metric induced
on the surfaces of constant $z$.
This metric evolves to a space-time with naked singularities on the
cone $\rho^2=t^2$. Near this cone the metric behaves like
\[
\sqrt{m}^{-1}e^{-U/2}\d z^2+e^{3U/2}\sqrt{m}(-\d T^2+\mbox{cosh}^2(T)\d\theta^2
+\d U^2), \qquad U\longrightarrow -\infty,
\]
where $\rho=e^U\ \mbox{cosh}(T), t=e^U\ \mbox{sinh}(T)$.
This solution represents a charged particle moving with the along  the $z$ axis. It can be interpreted as a 
tachyon in a sense of \cite{GR}, as it is unstable and invariant under $\R\times SO(2, 1)$. 

The properties of (\ref{example1}) signal the semi-classical instability
of the vacuum in Einstein--Maxwell--dilaton with attractive electric force.
The argument is analogous to Witten's bubble of nothing \cite{Wit}.
If the coordinate $z$ is periodic the solution is asymptotic to
the flat metric on $\R^3\times S^1$. 
The decay of $\R^3\times S^1$ vacuum 
is described by the instanton obtained by replacing $t\rightarrow 
i\tau$ in the metric (\ref{example1}). This instanton has vanishing action
\cite{GR}, and the probability of the decay is given by the exponential of
the negative action.

\section*{Acknowledgements}
This work resulted from a seminar \cite{gib_talk} given
at the programme Global Problems in Mathematical Relativity
held in Newton Institute  in the Fall 2005. I would like to thank  
Gary Gibbons and other participants of the programme for useful comments.

\end{document}